\input harvmac
\input epsf
\newcount\figno
\figno=0
\def\fig#1#2#3{
\par\begingroup\parindent=0pt\leftskip=1cm\rightskip=1cm
\parindent=0pt
\baselineskip=11pt
\global\advance\figno by 1
\midinsert
\epsfxsize=#3
\centerline{\epsfbox{#2}}
\vskip 12pt
{\bf Fig. \the\figno:} #1\par
\endinsert\endgroup\par
}
\def\figlabel#1{\xdef#1{\the\figno}}
\def\encadremath#1{\vbox{\hrule\hbox{\vrule\kern8pt\vbox{\kern8pt
\hbox{$\displaystyle #1$}\kern8pt}
\kern8pt\vrule}\hrule}}

\overfullrule=0pt

\Title{UK/02-01,TIFR-TH/02-02}
{\centerline{Thermality in de Sitter and Holography}} 
\smallskip
\centerline{Sumit R.  Das 
\foot{das@pa.uky.edu, das@theory.tifr.res.in}} 
\smallskip
\centerline{{\it Department of Physics and Astronomy,}}
\centerline{\it University of Kentucky,
Lexington, KY 40506, U.S.A.}
\smallskip
\centerline{and}
\smallskip
\centerline{{\it Tata Institute of Fundamental Research,}}
\centerline{\it Homi Bhabha Road, Bombay 400 005, INDIA.}

\bigskip

Assuming the existence of a $dS/CFT$ correspondence we study the
holograms of sources moving along geodesics in the bulk by calculating
the one point functions they induce in the boundary theory.  In
analogy with a similar study of uniformly accelerated sources in $AdS$
spacetime, we argue that comoving geodesic observers correspond to a
coordinate system on the boundary in which the one point function is
{\it constant}.  For $dS_3$ we show that the conformal transformations
on the boundary which achieve this - when continued suitably to
Lorentzian signature - induce nontrivial Bogoliubov transformations
between modes, leading to a thermal spectrum.  This may be regarded as
a holographic signature of thermality detected by bulk geodesic
observers.

\medskip

\noindent

\Date{January 2002}

\def\scri{{\cal I}}
\def\vx{{\vec x}}
\def\cO{{\cal O}}
\def\cQ{{\cal Q}}
\def\ba{{\bf a}}
\def\bb{{\bf b}}
\def\tbb{{\tilde{\bf b}}}
\def\bPhi{{\bf \Phi}}
\def\vk{{\vec k}}
\def\bu{{\bar u}}

\newsec{Introduction and summary}

There are several situations where quantum fields on curved
space-times lead to thermal behavior \ref\birell{See e.g. N.D. Birell
and P.C.W. Davies, {\it Quantum fields in curved space} (Cambridge
University Press, 1982)}.  The most dramatic example is Hawking
radiation from black holes \ref\hawking{S. Hawking, {\it Nature}
{\bf 248} (1974) 30; S. Hawking, {\it Comm. Math. Phys.} {\bf 43} (1975) 
199.}.
A second example is the thermal spectrum observed by uniformly
accelerated detectors in flat space (the Unruh effect)
\ref\unruh{W. Unruh, {\it Phys. Rev.} {\bf D14} (1976) 870.} - 
which is intimately related to black hole
radiation. Unruh radiation has generalizations to other space-times,
e.g. anti-de-Sitter (AdS) spacetimes \ref\deser{S. Deser and O. Levin,
{\it Class. Quant. Gr.} {\bf 14} L163; S. Deser and O. Levin,
{\it Class. Quant. Gr.} {\bf 15} (1998) L85, {\tt hep-th/9806223}; 
S. Deser and O. Levin,
{\it Phys. Rev.} {\bf D59} (1999) 064004, {\tt hep-th/9809159}
; T. Jacobson, {\it Class.Quant.Grav} {\bf 15} (1998)
251, {\tt  gr-qc/9709048}}.  Another set of
examples are cosmological
space-times \birell, the most well-studied instance being de-Sitter (dS)
spacetime. In this spacetime any geodesic observer perceives the
invariant vacuum as a thermal state \ref\desitter{G.W. Gibbons and 
S. Hawking, {\it Phys. Rev.} {\bf D15} (1977) 2738.}.

While recent developments in string theory have thrown valuable light
on the microscopic origin of black hole radiation
\ref\blackhole{A. Strominger and C. Vafa, {\it Phys. Lett.} {\bf
B379} (1996) 99, {\tt hep-th/9601029}; C. Callan and J. Maldacena
{\it Nucl. Phys.} {\bf B472} (1996) 591, {\tt hep-th/9602043}; 
A. Dhar, G. Mandal and S.R. Wadia, {\it Phys. Lett.} {\bf B388} (1996)
51, {\tt hep-th/9605234}  ;
S.R. Das and S.D. Mathur, {\it Nucl. Phys.} {\bf B478} (1996) 561,
{\tt hep-th/9606185 };
J. Maldacena and A. Strominger, {\it Phys. Rev.} {\bf D55} (1997) 861,
{\tt hep-th/9609026}.}, we know very little about thermal behavior in
cosmological spacetimes. This note is a attempt to throw some light on
this important issue for de Sitter spacetimes.

The microscopic origin of black hole thermodynamics led to a concrete
realization of the holographic principle \ref\holographic{
G. 't Hooft, in ``{\it Salamfest}'' (1993) 0284,
{\tt gr-qc/9310026}; L. Susskind, 
{\it J. Math. Phys.} {\bf 36} (1995) 6377, {\tt hep-th/9409089}.} - 
the $AdS/CFT$ correspondence \ref\adscft{J. Maldacena, {\it Adv. Theo. Math. Phys.} {\bf 2}
(1998) 231, {\tt hep-th/9711200}; S. Gubser, I. Klebanov and
       A. Polyakov, {\it Phys. Lett.} {\bf B428} (1998) 105, {\tt
       hep-th/ 9802109}; E. Witten, {\it Adv. Theo. Math. Phys.} {\bf
       2} (1998) 253, {\tt hep-th/9802150}.}.  
Conversely  $AdS/CFT$ duality has provided a physical
understanding of Hawking radiation and related phenomena in terms of
the dual field theory.  If the holographic principle is correct in this
context,
quantum gravity in de Sitter space-time should have a holographic dual
which is some theory living on one of the spacelike boundaries ${\cal
I}^\pm$. Our experience with $AdS/CFT$ duality suggests that
understanding this dual theory would throw valuable light on bulk
behavior in de Sitter space-times.

Unfortunately, we do not know how to obtain de Sitter space-time
from string theory in a fully satisfactory manner. In fact, under
some assumptions there appears to be a no-go theorem \ref\nogo{B. de Wit,
D. Smit and N,D, Hari Dass,
{\it Nucl.Phys.} {\bf B283} (1987) 165;J. Maldacena
Nucl.Phys.B283:165,1987 
and C. Nunez, {\it Int.J.Mod.Phys.} {\bf A16} (2001) 822.}, 
though there have been several proposals in the past as well as in
recent years
\ref\attempts{C. Hull, {\it JHEP} 9807 (1998) 021, {\tt hep-th/9806146}, 
E. Silverstein, hep-th/0106209; 
C.M. Hull, {\it JHEP 0111} (2001) 012, {\tt hep-th/0109213};
and {\it JHEP} {\bf 0111} (2001) 061, {\tt hep-th/0110048};
C. Madeiros, C. Hull and B. Spence, {\tt hep-th/0111190};
G. Gibbons and C. Hull, {\tt hep-th/0111072}.}
Nevertheless it is
important to figure out what would be the holographic signature of
bulk phenomena {\it assuming} that such a correspondence exists. In
this spirit \ref\balahorava{V. Balasubramanian, P. Horava and
D. Minic, JHEP {\bf 0105} (2001) 043, {\tt hep-th/0103171}}
\ref\witten{E. Witten, hep-th/0106109} 
\ref\strom{A. Strominger, hep-th/0106113} 
propose various versions of this correspondence, in direct analogy
with the $AdS/CFT$ correspondence. For other work in this direction,
see \ref\other{
M. Li, {\tt hep-th/0106184};
D. Klemm, {\tt hep-th/0106247};
S. Nojiri and S.D. Odinstov, {\tt hep-th/0106191, hep-th/0107134}; 
A. Tolley and N. Turok, {\tt hep-th/0108119};
T. Shiromizu, S. Ida and T. Torii, {\tt hep-th/0109057}; 
B. McInnes, {\tt hep-th/0110062};
A. Strominger, {\tt hep-th/0110087};
V. Balasubramanian, J. de Boer and D. Minic, {\tt hep-th/0110108};
B. Carniero de Cunha, {\tt hep-th/0110169};
R. Cai, Y. Myung and Y. Zhang, {\tt hep-th/0110234};
U. Danielsson, {\tt hep-th/0110265};
S. Ogushi, {\tt hep-th/0111008};
A. Petkou and G. Siopsis, {\tt hep-th/0111085};A.M. Ghezelbash and 
R.B. Mann, {\it JHEP} {\bf 0201} (2002) 005, {\tt hep-th/0111217};
A.M. Ghezelbash, D. Ida, R.B. Mann, T. Shiromizu,
{\tt hep-th/0201004}.}.
General aspects of holography in de Sitter spaces are discussed in
\ref\boussobanks{R. Bousso, hep-th/0012052, JHEP {\bf 9906} (1999) 028
[{\tt hep-th/9906022}] and JHEP {\bf 0011} (2000) 038
[{\tt hep-th/0010252}]; T. Banks, {\tt hep-th/0007146}}. 
While various aspects of these proposals
lead to interesting insights into the nature of the holographic
theory, a holographic explanation of the thermal behavior observed by
a geodesic observer is still a mystery. 

In this paper we throw some light on this issue.  We use an earlier work
\ref\dz{S.R. Das and A. Zelnikov, {\it Phys. Rev.} {\bf D64} (2001) 104001
[{\tt hep-th/0104198}]}, which addressed a similar question
in AdS spacetimes. In the latter situation, {\it uniformly
accelerated} observers measure a thermal spectrum, provided the
acceleration exceeds a critical bound \deser.  The purpose of \dz\ was
to ask : how does one understand this thermality in the holographic
theory ? To answer this question, \dz\ considered an external source
moving with a uniform acceleration. The source couples to one of the
supergravity fields in the bulk, e.g. the dilaton $\Phi$.  According
to the standard $AdS/CFT$ correspondence the value of the field $\Phi$
produced by this source is equal to the one point function of the
operator dual to this field in the boundary CFT
\ref\bulkboundary{V. Balasubramanian, P. Kraus and A. Lawrence,
{\it Phys. Rev.} {\bf D59} (1999) 046003, {\tt hep-th/9805171}; 
V. Balasubramanian, P. Kraus, A. Lawrence and
S. Trivedi, {\it Phys. Rev.} {\bf D59} (1999) 104021, 
{\tt hep-th/9808017}; T. Banks, G. Horowitz and
E. Martinec, {\tt hep-th/9808}; E. Keski-Vakkuri, 
{\it Phys. Rev.} {\bf D59} (1999) 104001, {\tt hep-th/9808037};
U. Danielsson, E. Keski-Vakkuri and M. Kruczenski,
{\it JHEP} {\bf 9901} (1999) 002, {\tt hep-th/9812007}.},
\ref\dasghosh{S.R. Das and B. Ghosh, {\bf JHEP} {\bf 0006} (2000) 043,
[{\tt hep-th/0005007}]}. This provides a
``hologram'' of the moving source. Consider such a source which moves
normal to the boundary, away from it. Generically, at some given time,
the hologram has a profile which is peaked at the point where the bulk
trajectory intersects the boundary, dying off away from it. The width
of this profile is related to the distance of the source from the
boundary in accordance with the IR/UV connection. Thus the profile is
time dependent, spreading as the source in the bulk moves deeper into
$AdS$ space. We now need to understand how to describe a {\it comoving} bulk
observer holographically.  This may be done by performing a conformal
coordinate transformation {\it on the boundary} such that the
transformed one point function is {\it time independent}. Such a
coordinate transformation on the boundary would generically mix up
positive and negative frequency modes of any field in the boundary
CFT. In \dz\ it was found that such mixing occurs only when the bulk
acceleration exceeds the critical value. In fact the metric of the
boundary generically now becomes time-dependent, i.e. a cosmological
space-time. Some of the cosmologies obtained in this way are well known.
The final
result is that there is a holographic relationship between
acceleration radiation in the bulk and thermal behavior in cosmologies
defined on the boundary.

In the following, we adopt the same strategy for geodesic trajectories
in de Sitter space. However, now there are crucial differences which
make the physical interpretation a bit confusing. The boundaries are
now space-like, $\scri^\pm$, and the dual theory is euclidean. In
planar coordinates, time evolution in the bulk maps into {\it
decreasing} radial distance on the boundary.
Furthermore, unlike the $AdS/CFT$ correspondence it is not yet clear
whether there is an operator correspondence in $dS/CFT$. In fact, as
we shall see below, an operator correspondence is not necessarily
equivalent to a correspondence between the bulk effective action and
the CFT free energy in the presence of sources. Nevertheless, in this
work we will assume an operator correspondence. Then one point
functions of dual operators are related to the value of the bulk field
on the boundary - apart from the standard factor involving a UV
cutoff.
 
We consider a source for some scalar field $\Phi$ of mass $m < 1$ (in
units where the de Sitter scale is set to unity) moving along a
geodesic in three dimensional de Sitter space.  We calculate the value
of the field on a cutoff boundary, and hence the one point function of
the dual operator, on the boundary $\scri^+$. While the field $\Phi$
is a scalar, the definition of the cutoff boundary, and hence the one
point function, depends on the coordinate system used. In planar
coordinates, the one point function peaks at the point where the
geodesic intersects $\scri^+$ and decays as a power of the radial
distance.  In analogy with \dz\ we then ask whether there is a
coordinate transformation {\it on the boundary} which renders the
transformed one point function {\it constant} over the entire
boundary. This is indeed possible since the dual operator has
nontrivial conformal dimensions. Boundary (euclidean) correlation
functions which are single valued in planar coordinates are {\it
periodic} in one of the new coordinates, and can therefore be
interpreted as thermal Green's functions of a Lorentzian signature
theory.  After an analytic continuation to Lorentzian signature, this
new coordinate system is the holographic interpretation of a comoving
observer along a geodesic. The coordinate transformation, when
analytically continued in this fashion, mixes positive and negative
frequency modes of fields in the boundary theory - leading to the
correct temperature. An entirely analogous story holds for holograms
in global coordinates as well.

The coordinate transformation involved is once again a conformal
transformation and is in fact the restriction of the bulk
transformation which takes the planar or global coordinates to
``static'' coordinates to the boundary. In fact, the field produced on a
cutoff boundary defined in terms of a ``static'' coordinate system is
constant. As emphasized above, while the field $\Phi$ is a scalar, the
one point function transforms nontrivially as a conformal field -
which explains the result. The final offshoot is that thermality in
the dual description appears because of a nontrivial Bogoliubov
transformation involved in the passage to the natural holographic analogs
of ``comoving'' bulk observers.  

While the above results pertain to $dS_3$ we believe that the picture
is similar for other $dS_d$, though we do not present explicit
computations.

In this work we have used retarded Green's functions to determine the
field due to the geodesic source which is then identified with the
one point function in the boundary theory. While the rationale for this
is clear in the AdS/CFT correspondence, it is not so in the present
case. We comment, without explicit calculations, the relevance of
this issue to recent results which concern the behavior of geodesic
detectors in the bulk in the one parameter class of de Sitter invariant
states.

Section 2 contains definitions of various coordinate systems used in
this work. In Section 3 we discuss an operator version of the dS/CFT
correspondence. Section 4 contains the calculation of the hologram
of a geodesic source in planar coordinates. Section 5 discusses
interpretations of the hologram in both planar and global coordinates
and possible extensions to $dS_d$ for $d \neq 3$. Section 6 contains
conclusions and comments.

\newsec{Coordinate systems in de Sitter}

Throughout the paper, all dimensional quantities are measured in 
units of the de Sitter length scale.

$d+1$ dimensional de-Sitter space is a hyerboloid in $d+2$ dimensional
flat space with signature $(-1,1,1 \cdots 1)$ defined by the equation
\eqn\aone{-(Y^0)^2 + (Y^1)^2 + \cdots (Y^{d+1})^2 = 1}
Various coordinate systems are given by different ways of solving this
equation.

Global coordinates are defined by the embedding
\eqn\atwo{\eqalign{ Y^0 & = \tan T \cr
 Y^1 & = \sec T ~\cos \theta_1 \cr
Y^2 & = \sec T ~\sin \theta_1 \cos \theta_2 \cr
& \cdots \cr
Y^{d+1} & = \sec T~\sin \theta_1 \cdots \sin \theta_d}}
where
\eqn\athree{\eqalign{& -{\pi\over 2} \leq T \leq {\pi \over 2} \cr
& 0 \leq \theta_i \leq \pi ~~~~~~i = (1, \cdots (d-1))\cr
& 0 \leq \theta_d \leq 2\pi}}
The de-Sitter metric is then
\eqn\athree{ds^2 = \sec^2 T[-dT^2 + d\Omega_d^2]}
where $d\Omega_d^2$ is the metric on a unit $S^d$ whose coordinates are
$\theta_1 \cdots \theta_d$. The penrose diagram may be drawn using 
\athree\ directly, as shown in Fig. 1. The future infinity $\scri^+$ is
at $T = {\pi \over 2}$ while the past infinity $\scri^-$ is at 
$T = -{\pi \over 2}$ The diagram supresses the angles $\theta_2 \cdots
\theta_{d+1}$ and north pole corresponds to $\theta_1 =0 $ while the
south pole is at $\theta_1 = \pi$.

\fig{Penrose diagram of de Sitter space. The curved line is a constant
${\hat t}$ surface.}
{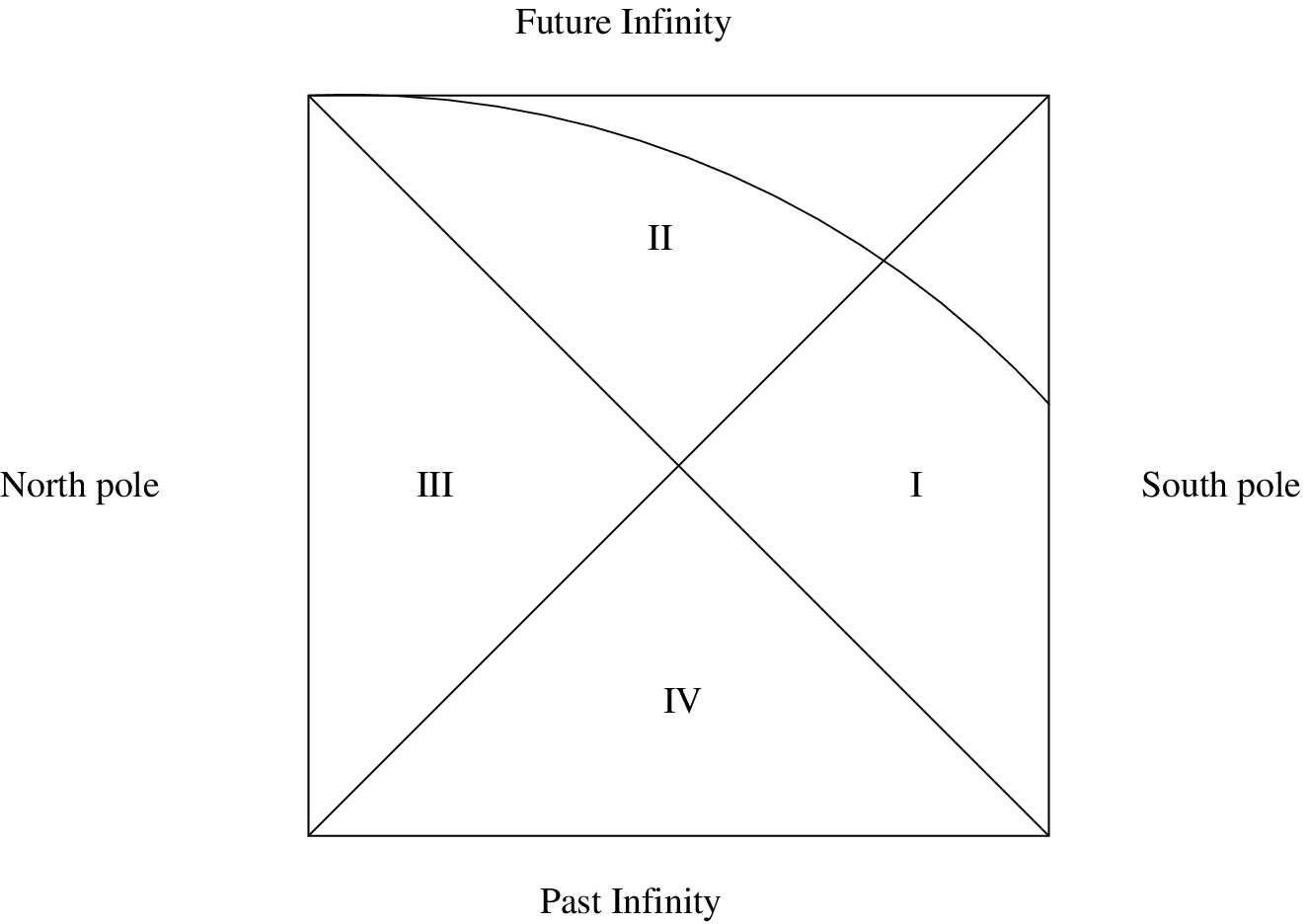}{3.8in}

Planar or steady-state coordinates which cover regions I and II in Fig. 1
are defined by
\eqn\afour{\eqalign{ Y^0 & = {1\over 2} e^{{\hat t}} \rho^2 + \sinh {\hat t} \cr
Y^1 & = {1\over 2} e^{{\hat t}} \rho^2 - \cosh {\hat t} \cr
Y^2 & = e^{\hat t} \rho \cos \theta_2 \cr
Y^3 & = e^{\hat t} \rho \sin \theta_2 \cos \theta_3 \cr
& \cdots \cr
Y^{d+1} & = e^{\hat t} \rho \sin \theta_2 \cdots \sin \theta_{d}}}
The angles $\theta_i$ in \afour\ are the same as in \atwo. $\rho$ is a
radial coordinate in $R^d$, $0 \leq \rho \leq \infty$ 
which is formed by $\rho$ and the $d-1$ angles
$\theta_2 \cdots \theta_d$. The metric now reads
\eqn\afive{ ds^2 = - d{\hat t}^2 + e^{2{\hat t}}(d\rho^2 + 
\rho^2 d\Omega_{d-1}^2)}
It is sometimes convenient to introduce cartesian coordinates on the $R^d$
which we denote by $x^i$, and also introduce a time coordinate $y$
\eqn\afiveb{ y = e^{\hat t}}
in terms of these the metric becomes
\eqn\afivea{ds^2 = {1\over y^2}[-dy^2 + dx^i dx^j \delta_{ij}]}
Comparing \atwo\ and \afour\ it is easy to see that this coordinate system
covers only regions I and II. This is because \afour\ implies that
$Y^0 - Y^1 = e^{{\hat t}} > 0$, while from \atwo\ we get
\eqn\asix{Y^0 - Y^1 = 2 \sec T~\cos [{1\over 2}(T+{\pi \over 2}-\theta_1)]
\sin [{1\over 2}(T-{\pi \over 2}+\theta_1)]}
Since both $[{1\over 2}(T+{\pi \over 2}-\theta_1)]$ and 
$[{1\over 2}(T-{\pi \over 2}+\theta_1)]$ range from $-{\pi \over 2}$ to
${\pi\over 2}$, the first two factors in \asix\ are always positive, so that
the sign is determined by the sign of $[{1\over 2}(T-{\pi \over 2}+\theta_1)]$.
The latter is positive in regions I and II. Planar coordinates which cover
regions III and IV may be defined in an analogous fashion.

A third coordinate system will be called ``static'' coordinates. In Region
I we have
\eqn\aseven{\eqalign{Y^0 & = {\sqrt{1-r^2}}~\sinh t \cr
Y^1 & = -{\sqrt{1-r^2}}~\cosh t \cr
Y^2 & = r~ \cos \theta_2 ~~~~~~~~~~~~~~~~~~~{\rm Region~I}\cr
Y^3 & = r~  \sin \theta_2 \cos \theta_3 \cr
& \cdots \cr
Y^{d+1} & = r~\sin \theta_2 \cdots \sin \theta_{d}}}
where $-\infty \leq t \leq \infty$ and $0 \leq r \leq 1$. The angles
$\theta_2 \cdots \theta_{d-1}$ are the same in \atwo\ and \afour.
The metric is
\eqn\aseven{ds^2 = -(1-r^2) dt^2 + {dr^2 \over 1-r^2} + r^2 d\Omega_{d-1}^2}
The south pole
is given by $ r = 0$ while the past and future horizons are given by 
$r = 1$. The metric is time independent in this region, which is why
these are called static coordinates.

In region II we have and $1 \leq r \leq \infty$
\eqn\aeight{\eqalign{Y^0 & = {\sqrt{r^2-1}}~\sinh t \cr
Y^1 & = -{\sqrt{r^2-1}}~\cosh t \cr
Y^2 & = r~ \cos \theta_2 ~~~~~~~~~~~~~~~~~~~{\rm Region~II}\cr
Y^3 & = r~  \sin \theta_2 \cos \theta_3 \cr
& \cdots \cr
Y^{d+1} & = r~\sin \theta_2 \cdots \sin \theta_{d}}}
The metric is again given by \aseven, but $r$ rather than $t$ is a timelike
coordinate in this region. Thus the metric is not stationary any more.
However we will retain the nomenclature ``static coordinates'' even in this
region.
The future infinity $\scri^+$ is given by $r = \infty$.
It is possible to introduce Kruskal coordinates which cover the entire
space-time. These are denoted by $(U,V, \theta_2 \cdots \theta_d)$ where
\eqn\anine{\eqalign{& r = {1+UV \over 1-UV}~~~~~~
t ={1\over 2} {\rm log}~(-{U\over V}) ~~~~~~{\rm Region~I}\cr
& r = {1+UV \over 1-UV}~~~~~~
t ={1\over 2} {\rm log}~({U\over V}) ~~~~~~{\rm Region~II}}}
Thus in region I $U > 0, V < 0$ while in region II $U,V > 0$.
In terms of these Kruskal coordinates we have in both regions I and II
\eqn\aten{\eqalign{Y^0 & = {U+V \over 1-UV} \cr
Y^1 & = {V-U \over 1-UV} \cr
Y^2 & = {1+UV \over 1-UV} \cos \theta_2 \cr
Y^3 & ={1+UV \over 1-UV} ~  \sin \theta_2 \cos \theta_3 \cr
& \cdots \cr
Y^{d+1} & = {1+UV \over 1-UV}~\sin \theta_2 \cdots \sin \theta_{d}}}

\newsec{An operator $dS/CFT$ correspondence}

According to the $dS/CFT$ correspondence, physics in the bulk of $dS_d$ has
a holographic dual which is a conformal field theory living on the boundary
of the space-time. In global coordinates the boundaries could be either
$\scri^+$ or $\scri^-$, but not both \strom. 
For bulk fields which satisfy standard wave equations with two
derivatives, the boundary data are the values of the field 
on $\scri^+$ and $\scri^-$. Equivalently, one considers the two
independent solutions of the equations of motion and specifies the
data in terms of these, as in \strom. From the point of view of
a formulation of the $dS/CFT$ correspondence which relates
the bulk effective action to the free energy
of the CFT in the presence of sources, {\it both} the solutions
have to be retained - in contrast to the $AdS/CFT$ correspondence. As 
a result there are {\it two} dual CFT operators for each bulk field.

In planar coordinates which cover regions I and II, and static
coordinates which cover region II, the boundary is at
$\scri^+$. However, since these coordinates do not cover the entire
space-time, one has to specify the values of various bulk fields along
the horizons. Equivalently one has to again consider {\it both} the
independent solutions of the equations of motion.

In the $AdS/CFT$ correspondence there is an operator formulation
\bulkboundary. We assume that there is a similar operator version of the
$dS/CFT$ correspondence.  We will spell out this operator
correspondence in planar coordinates, though similar considerations
are valid for global coordinates as well.

Consider a massive scalar field $\Phi$ of mass $m$ in $dS_{d+1}$. We will
consider the case $m < d/2$ in $dS$ units.
The free Klein-Gordon equation is
\eqn\bone{ (\nabla^2 - m^2) \Phi = 0}
In planar coordinates given by \afivea, the two independent solutions may
be easily written down 
\eqn\btwo{\eqalign{\Phi^{(1)}_k ({\vec x},y) 
& = {1\over 2 (2\pi)^{d-1\over 2}} 
y^{d/2}~H_\nu^{(1)}(|k|y)~e^{i \vk \cdot \vx} \cr
\Phi^{(2)}_k ({\vec x},y) & = {1\over 2 (2\pi)^{d-1\over 2}}
y^{d/2}~H_\nu^{(2)}(|k|y)~e^{i \vk \cdot \vx}}}
where $H_\nu^{(i)}$ are Hankel functions and
\eqn\bthree{\nu = +{\sqrt{(d/2)^2-m^2}}}

The modes have been chosen so that they are complex conjugates of each
other and normalized according to the standard Klein Gordon inner
product. Therefore the mode expansion which defines the
creation/annihilation operators is
\eqn\bfour{\Phi({\vec x},y) = \int {d^d k \over (2\pi)^{d}}[
\Phi_k^{(1)} ({\vec x},y)~ {\bf a} (\vk) + {\rm h.c.}]}
The operators $\ba (k)$ satisfy
\eqn\bfoura{ [ \ba (\vk), \ba^\dagger (\vk ') ] = \delta^d (\vk - \vk ')} 
The Fock vaccuum in these coordinates is then defined by
\eqn\bfive{ {\bf a} (\vk) |0> = 0~~~~~~~{\rm for~all}~\vk}
and the states are labelled as usual by the values of the momenta $\vk$. 
Single
particle states are
\eqn\bsix{ |\vk> = {\bf a}^\dagger (\vk) |0>}
Consider the mode expansion \bfour\ close to $\scri^+$, i.e. $y = y_0 
\rightarrow 0$.
The leading order result for the field operator is then, upto 
numerical factors
\eqn\bseven{\Phi (\vx, y_0) \sim {y_0^{{d\over 2}-\nu}\over i\Gamma (1-\nu)
\sin~\pi\nu}\int d^d k~k^\nu~[\ba (\vk) - \ba^\dagger (-\vk)]
e^{i\vk \cdot \vx}}
where $k \equiv |\vk|$.
Thus we can define a boundary operator $\cO_- (k)$ by
\eqn\beight{ \cO_-(\vk) \equiv k^\nu[\ba (\vk) - \ba^\dagger (-\vk)]}
which is the fourier transform of some local operator $\cO_-(\vx)$ on the
boundary. The power of $y_0$ clearly indicates that the conformal dimension
of this operator is 
\eqn\bnine{\Delta_- = {d\over 2} - \nu}
This is an operator form of $dS/CFT$ correspondence. 

Note that roughly {\it half} of the bulk operators are related to the
specific boundary operator which arises from restricting the field to
the boundary. The Hankel functions have two pieces $J_\nu (|k|y)$ and
$J_{-\nu}(|k|y)$. Near $\scri^+$ the latter dominates and the operator
which comes with it is what we have identified above. There is another
operator $\cO_+ (k)$, which comes with the other Bessel function
$J_\nu (|k|y)$ and is independent of $\cO_-(k)$.  In position space
one has
\eqn\bninea{{\rm Lim}_{y_0 \rightarrow 0}~\Phi (y_0, \vx)
\sim (y_0)^{d/2-\nu} \cO_- (\vx) + \sim (y_0)^{d/2+\nu} \cO_+ (\vx)}
This is the operator
manifestation of the appearance of {\it two} dual operators for a
single bulk field, as observed in \strom.

This is in sharp contrast with the situation in $AdS/CFT$ correspondence.
In Poincare coordinates of $AdS_{d+1}$
\eqn\bten{ds^2 = {1\over z^2} [-dt^2 + dz^2 + d \vx \cdot d \vx]}
the mode expansion of a massive scalar field is
\eqn\beleven{\Phi (z,x,t) \sim z^{d/2}\int_0^\infty d\alpha\int d^{d-1}k~
({\alpha \over \omega})^{1/2}~J_\mu (\alpha z)[\bb (\vk,\alpha) e^{-i(\omega
t - \vk \cdot \vx)} + h.c.]}
where 
\eqn\btwelve{\omega^2 = \vk^2 + \alpha^2~~~~~~~\mu= +{\sqrt{(d/2)^2+m^2}}}
The mode expansion involves only one of the independent solutions of the
Klein Gordon expansion since the other solution is not normalizable.
In this case, the field operator near the boundary $z = z_0 \rightarrow 0$
gets indentified with a local boundary operator $\cQ (x,t)$ by
\bulkboundary, \dasghosh\
\eqn\bthirteen{ {\rm Lim}_{z_0 \rightarrow 0}~\Phi (z_0,x,t) =
(z_0)^{\mu + d/2}~\cQ(x,t)}
where the fourier transform of $\cQ(x,t)$ on the boundary, $\cQ (k,\omega)$
is given in terms of the annihilation and creation operators in 
\beleven by \dasghosh\
\eqn\bfourteen{\cQ (\omega, \vk) \sim (\omega^2 - k^2)^{\mu / 2}[
\theta (\omega)~\tbb (\omega, \vk) + 
\theta (-\omega)~\tbb^\dagger (-\omega, - \vk)]}
where
\eqn\bfourteena{\tbb (\omega, \vk) = ({\omega \over \alpha})^{1/2} 
\bb (\alpha, \vk)}
In this case {\it all} of the annihilation and creation operators of the bulk
field are necessary to construct the boundary operator. Consequently, for
a given bulk field there is only one dual operator.

In the $AdS/CFT$ correspondence, the bulk and the boundary share the
same time. If there is a $dS/CFT$ correspondence, the dual theory is
euclidean. From the form of the metric it is clear that rescaling time
is equivalent to rescaling distances on $\scri^+$. In the above correspondence
we have used the momenta $\vk$ to label states of the CFT. This means
we are considering one of the coordinates on $\scri^+$, $x_2$ or $x_3$
as the euclidean time. A proper interpretation would however involve
a continuation to lorentzian signature on the boundary.

Assuming that there is such a $dS/CFT$ correspondence, it is clear
that the correlation functions of boundary operators can be written in
terms of the Green's functions of the bulk fields. Such an assumption
has been made in e.g. \ref\bms{R. Bousso, A. Maloney and
A. Strominger, {\tt hep-th/ 0112218}} and \ref\sv{M. Spradlin and
A. Volovich, {\tt hep-th/0112223}}.  However it is not clear how
this is related to a $dS/CFT$ correspondence based on the equality of
the effective action of the bulk theory and conformal field theory
free energy in the presence of sources. For example, \bms,\sv
show that an operator correspondence leads to different results for
CFT correlators for different members of the one parameter family of
de Sitter invariant vacua found in \ref\mot{E. Mottola, {\it
Phys. Rev.} {\bf D 31} (1985) 754; B. Allen, {\it Phys. Rev.} {\bf D
32} (1985) 3136.}. However if these CFT correlators are calculated
according to the prescription of \strom, they are independent of the
value of the parameter. This is because the Green's functions
(satisfying the inhomogeneous equation) for different values of this
vacuum parameter differ from one another by solutions of the
homogeneous equation and this addition does not change the field
evaluated on the boundary according to the procedure of
\ref\giddings{S.  Giddings, {\it Phys. Rev. Lett.} {\bf 83} (1999)
2707 [{\tt hep-th/9903048}]}.

In the following we will assume an operator correspondence. While we
have described this in the planar coordinate system, it is clear that
this can be done in global coordinates using the modes derived in
\ref\dsmodes{N.A. Chernikov and E.A. Tagirov, {\it Ann. Poinc. Ohys. Theor.} 
{\bf A 9} (1968) 109; E.A. Tagirov, {\it Annals. Phys.} {\bf 76} (1973) 561;
R. Figari, R. Hoegh-Krohn and C. Nappi, {\it Comm. Math. Phys.} {\bf 44} 
(1975) 265; H. Rumpf and H.K. Urbantke, {\it Ann. Phys.} {\bf 114} (1978)
332; L. Abbott and S. Deser, {\it Nucl. Phys.} {\bf B 195} (1982) 76;
A.H. Hajmi and A. Ottewill, {\it Phys. Rev} {\bf D 30} (1984) 1733;
L. Ford, {\it Phys. Rev.} {\bf D 31} (1985) 710; B. Allen and A. Folacci,
{\it Phys. Rev.} {\bf D 35} (1987) 3331; C.J. Burgess, {\it Nucl. Phys.}
{\bf B 247} (1984) 533.} and \mot, or for any other coordinate system.

\newsec{Holograms of Geodesics in $dS_3$}

The simplest geodesic trajectory is the worldline of the south
pole. In planar coordinates this is described by $\rho = 0$ while in
static coordinates in region I this corresponds to $r = 0$. All other
geodesics are obtained from this by isometries of de Sitter space. It
is clear that the trajectory of any point of the spatial $S^d$ in
global coordinates is a geodesic.  Similarly any point on the $R^d$ in
planar coordinates is a geodesic as well.  Because of the maximal
symmetry of the space it is sufficient to consider the geodesic at the
south pole.

Consider a source moving along the geodesic in $dS_3$ which couples to
a bulk scalar field $\Phi$ of mass $m$. We work in planar coordinates which
cover regions I and II of the Penrose diagram. 
According to the previous section, the leading value \foot{We work in
the leading order of the semiclassical expansion of the bulk theory.} of the
one point function of the dual operator $\cO_- (\vx)$ in this state is
given in terms of the value of the scalar field by
\eqn\cone{<\cO_- (\vx) >
\sim {\rm Lim}_{y_0 \rightarrow}~(y_0)^{1-\nu}~\Phi (y_0, \vx)}
The field $\Phi (y, \vx)$ is produced by the source. 
When the source is at $(y,\vx) = (y'(\lambda),0)$ where $\lambda$ is the
proper time along the geodesic, this is given by
\eqn\ctwo{\Phi (y, \vx) = \int d\lambda~ G_R (y,\vx; y'(\lambda),0)}
where $G_R(y,\vx;y',\vx')$ is the retarded Green's function with $y < y'$.
The latter is given by 
\eqn\cthree{G_R(y,\vx;y',\vx ') = i \theta (y' - y)~
<0| [ \bPhi (y,\vx), \bPhi (y',\vx ')]|0>}
where $\bPhi$ denotes the field operator. This may be readily calculated 
using the mode expansion \bfour. For $d = 3$ one has $\nu = {\sqrt{1-m^2}}$
so that $\nu < 1$ and non-integral. We can then use the expressions for
the Hankel functions
\eqn\cfour{\eqalign{H^{(1)}_\nu(z) & = {1\over i~\sin\pi\nu}[J_{-\nu}(z)
- e^{-i\pi\nu} J_\nu (z)]\cr
H^{(2)}(z) & = {1\over i~\sin\pi\nu}[e^{i\pi\nu} J_{\nu}(z)
- J_\nu (z)]}}
The result for $G_R$ is then
\eqn\cfive{G_R(y,\vx;y',\vx ')= {2\theta (y' - y)\over~
\sin \pi\nu}\int {d^2 k \over 16 \pi^4} (y y')
e^{i \vk \cdot (\vx - \vx')}[J_{-\nu}(|k|y)J_{\nu}(|k|y') -
J_{-\nu}(|k|y')J_{\nu}(|k|y)]}
Along the trajectory $\vx = 0$ the proper time interval $d\lambda$ is
related to the increment in the coordinate time $dy'$ by
\eqn\csix{ d\lambda = - {dy' \over y'}}
Combining \ctwo, \cfive\ and \csix\ we finally get
\eqn\cseven{ \Phi (y,\vx) \sim
\int_\infty^y {dy' \over y'}\int d^2 k ~(y y')e^{i\vk \cdot \vx}
[J_{-\nu}(|k|y)J_{\nu}(|k|y') - J_{-\nu}(|k|y')J_{\nu}(|k|y)]}
where we have ignored inessential constants.
The field $\Phi$ is of course a scalar. Therefore the expression in any
other coordinate system may be obtained by simply reexpressing the right
hand side of \cseven\ in terms of the new coordinates. Alternatively one
can start out with a mode decomposition in the new coordinates.

To extract the hologram, i.e. the one point function of the dual operator
we have to take a limit of \cseven\ when $y = y_0 \rightarrow 0$. In this 
limit the dominant contribution comes from the term $J_{-\nu}(|k|y)
J_{\nu}(|k|y')$ in \cseven, which behaves as $|k|^{-\nu}(y_0)^{-\nu}
J_{\nu}(|k|y')$. Peforming the angular integral in momentum space we get
the result
\eqn\ceight{{\rm Lim}_{y_0 \rightarrow 0}~\Phi(y_0,\vx)
= (y_0)^{1-\nu}\int_\infty^{y_0} dy''\int_0^\infty d|k|~|k|^{1-\nu}~J_0(|k|\rho)
J_\nu(|k|y'')}
where $\rho = |\vx|$ as before.
The integral over $y'$ may be performed using
\eqn\cnine{\int_\infty^{y_0} dy'~ J_\nu (|k| y')
= {1\over |k|} [-1 + 2 \sum_{n=0}^\infty J_{2n+1-\nu}(|k|y_0)]}
Since we have $\nu < 1$ we get
\eqn\cten{{\rm Lim}_{y_0 \rightarrow 0}~\Phi(y_0,\vx)
= (y_0)^{1-\nu}\int_0^\infty d|k|~|k|^{-\nu}~J_0(|k|\rho)
\sim ( {y_0 \over \rho} ) ^{1-\nu}}
which leads to a one point function
\eqn\celeven{< \cO_- (\vx)>_{planar} \sim {1\over \rho^{1-\nu}}}
The power is appropriate for that of an operator with dimensions
$(\half (1-\nu), \half (1-\nu))$ in the two dimensional euclidean CFT
on the boundary.

\newsec{Interpreting the hologram}

In the bulk, an observer comoving with the geodesic will perceive the
vacuum as a thermal state with a temperature $T = {1/2\pi}$ in our
units. We want to see how is this reflected in the holographic
dual.

\subsec{Accelerated objects in $AdS_3$ (Poincare boundary)}

Before we start to interpret the hologram of a geodesic source in $dS_3$ let
us recall the main results of a similar calculation of holograms of
{\it accelerated} objects in $AdS_3$ coupled to a bulk massless scalar field
\dz. Consider the following trajectory
\eqn\done{t = \alpha ~z}
in a Poincare coordinate system
\eqn\dones{ds^2 = {1\over z^2}[-dt^2 +dz^2 +dx^2]}
This has a uniform invariant acceleration $a$ given by
\eqn\dtwo{a^2 = {\alpha^2 \over \alpha^2 -1}}
The parameter $\alpha$ labels the specific trajectory. A set of observers
comoving with this class of objects records a Unruh temperature
$T_U = {1\over 2\pi}$. The local temperature measured by a particular 
trajectory is related to $T_U$ by a redshift factor, leading to
$T = {1\over 2\pi{\sqrt{\alpha^2-1}}}$

When such an accelerated object couples to a massless scalar field in the
bulk, the one point function of the dual operator in the boundary CFT defined
on a cutoff boundary at $z=z_B \rightarrow 0$ can
be calculated along the lines of the previous section, with the result
\dz\
\eqn\dthree{<\cO> \sim \alpha {\sqrt{\alpha^2 -1}}{t \over [(\alpha^2-1)
x^2 + t^2]^{3/2}}}
As the Poincare time $t$ increases the object moves deeper into the
bulk of $AdS$ spacetime. The hologram, given by \dthree, reflects this :
the support of the one point function spreads with time.

Consider an observer {\it on the boundary} according to whom the
one point function is {\it time-independent}. This would be the hologram of
a bulk observer co-moving with the accelerated object. To find such observers
one must therefore look for a new coordinate system in which the one point
function is time independent. It turns out that this is in fact a 
conformal transformation on the boundary. Recalling the fact that the
conformal dimensions of $\cO$ are $(1,1)$ it is easy to see that 
the new time $\eta$ and the new
space $\psi$ are related to the original coordinates $(t,x)$ by
\eqn\dfour{t \pm x = {1\over \beta}e^{-\beta(\eta \mp \psi)}}
where $\beta$ is determined by requiring that the proper time interval
for any section of the trajectory is equal to the interval in terms of the
new time $\eta$,
\eqn\dfive{\beta = {1\over {\sqrt{\alpha^2-1}}}}
Then the transformed one point function is
\eqn\dsix{ <\cO_{\eta,\psi}> \sim \alpha {\sqrt{\alpha^2 -1}}
{\cosh \beta \psi \over [\alpha^2 \sinh^2 \beta\psi + 1]^{3/2}}}

The transformation \dfour\ covers only one wedge of the full Minkowski
space $(t,x)$ - the wedge which corresponds to the future light cone of
the point $t=x=0$. The original boundary metric becomes
\eqn\dseven{e^{-2\beta \eta}[-d\eta^2 + d\psi^2]}
This is the metric of a Milne universe. It is well known that the 
Minkowski vacuum appears as a thermal state in terms of particles defined
according to postive frequency using the time $\eta$, with a temperature
$T = {\beta \over 2\pi}$ \birell, \ref\milne{T. Tanaka and M. Sasaki,
{\it Phys. Rev.} {\bf D55} (1997) 6061.}
. This is exactly the bulk temperature.
The upshot is that acceleration radiation in the bulk is interpreted as
a {\it cosmological} radiation in the boundary theory.

There is in fact a good reason why \dfour\ is the correct transformation.
To see this we consider a different coordinate system in $AdS_3$, which
we call BTZ coordinates. The spatial coordinates are $\rho$ with a
range $1 < \rho < \infty$ and $\psi$ with range $0 < \psi < \infty$
while the time coordinate $\eta$ has a range $-\infty < \eta < +\infty$.
The metric now reads
\eqn\dseven{ds^2 = - (\rho^2 -1) d\eta^2 + {d\rho^2 \over \rho^2 -1}
+ \rho^2 d\psi^2}
The boundary is now at $\rho = \infty$, and the boundary coordinates on
a cutoff boundary $\rho = \rho_B$ are $(\eta,\psi)$. In these coordinates
the accelerated trajectory \done\ simply corresponds to
\eqn\deight{\rho = \alpha}
It is clear that if we calculate the one point function on a cutoff
boundary using these coordinates the result will be independent of the
time $\eta$. 

This explains why \dfour is the correct transformation. 
The point is that the transformations \dfour\
are precisely the coordinate transformation between
the Poincare coordinates $(t,z,x)$ and BTZ coordinates $(\eta,\rho,\psi)$
when restricted to a cutoff boundary at $z = z_0$or equivalently $\rho 
= {1\over z_0}$.

It is important to realize that while $\Phi$ is scalar under coordinate
transformations, the defintion of the cutoff boundary depends on the
specific coordinate system used \foot{This feature has been also observed
in computations of the Casimir energy, see Ghezelbash et.al. in last
reference in \other.}. This makes the one point function
coordinate dependent - in fact it just transforms as a conformal field
with the appropriate conformal dimension.

The treatment for other coordinates in $AdS$ is entirely analogous and
has been discussed in detail in \dz.

\subsec{Geodesics in $dS_3$ : Planar boundary}

The dual theory for $dS_3$ is euclidean. It is clear from the form
of the metric that in planar coordinates time evolution in the bulk
maps into scale evolution on the boundary.
In terms of the complex coordinate
\eqn\eone{z = x^2 + i x^3 = \rho e^{i\theta_2}}
the scale is represented by $\rho$.

In analogy with the case of accelerated objects in $AdS_3$ we therefore
ask : is there a conformal transformation
\eqn\etwo{ z \rightarrow w = w(z)}
which renders the one point
function \celeven\ independent of $|w|$ ? Because of the symmetry of the
problem this means that the transformed one point function is in fact
a constant.

This is indeed possible.  Using the fact that the operator $\cO_-$ has
dimensions $({\Delta_- \over 2},{ \Delta_-\over 2})$ where $\Delta_- =
1 - \nu$ it is easy to see that the transformation is
\eqn\ethree{z = e^w}
Our discussion of a possible operator correspondence in planar
coordinates imply that we can label the states of this euclidean
theory by the spatial momenta of the bulk theory. In other words, one
of the coordinates $x^2$ or $x^3$, e.g. $x^3$ can be regarded as an
euclidean time. Analytic continuation in $x^3$ then provides one definition
of a quatum theory on the boundary.

Upon analytic continuation $x^3 \rightarrow i x^3$
the one point function \celeven\ becomes
\eqn\eeone{< \cO_- (\vx)>_{planar} \sim {1\over [(x^2)^2 - (x^3)^2]^
{({1-\nu \over 2 })}}}
Now make the transformation to coordinates $(\sigma_1,\sigma_2)$
\eqn\eetwo{ x^2-x^3 = -e^{-(\sigma_2-\sigma_3)}~~~~~
x^2+x^3 = e^{-(\sigma_2+\sigma_3)}}
The transformed one point function is then independent of $\sigma_i$.
The transformation \eetwo\ is precisely the transformation between
Minkowski and Rindler coordinates in two dimensional flat space. As is well
known this induces a nontrivial Bogoliubov transformation between modes
and the Miknowski vacuum appears as a thermal state in terms of Rindler
particles with the temperature given by 
\eqn\efour{T = {1\over 2\pi}.}

For the special case ($AdS_3$) we are considering, there is another way to
understand this. Define
\eqn\ethreeb{ w = \sigma_1 + i \sigma_2}
It is then clear that correlation functions which are single valued in
$(x^2,x^3)$ would be periodic in $\sigma_2$. This means that in an
alternative definition of a quantum theory on the boundary in which
$\sigma_2$ is regarded as an euclidean time, these correlators would
be {\it thermal} with a temperature given by \efour.
This is exactly the temperature measured by a geodesic observer in the
bulk. Note this is not the usual way a field theory on a cylinder is
defined. As we will see later, this argument needs
a modification in
higher dimensions.

Once again there is a good reason why this is the right conformal
transformation. The geodesic $\vx = 0$ corresponds to the point
$r = 0$ in the static coordinate system in the patch I. This coordinate
patch does not contain $\scri^+$, but the coordinate $r$ can be continued
to the coordinate $r$ in the static patch in region II containing $\scri^+$.
Our experience with $AdS$ leads us to expect that the one point function
in these coordinates is in fact a constant. 

Let us check this explicitly.  Since $\Phi$ is a scalar field, all we
have to do is to find the coordinate transformation relating the
static and planar coordinates and express the expression for $\Phi$ at
some general point, equation \cseven, in terms of static
coordinates. It is important that we do this before we take any limit
which takes us to a boundary since cutoff boundaries in different
coordinate systems do not coincide.

The coordinate transformations can be read off from \afour,\aseven,
\aeight\ and \aten. In Kruskal coordinates we have
\eqn\efive{ \rho = {1 + UV \over 2U}~~~~~~~y = {1-UV \over 2U}}
in both regions I and II, while in terms of the $(r,t)$ coordinates
we have in region I
\eqn\esix{ \rho = {r \over {\sqrt{1-r^2}}}~e^{-t}~~~~~~~~
y = {1 \over {\sqrt{1-r^2}}}~e^{-t}~~~~~~~~({\rm Region~I})}
while in region II
\eqn\esixa{ \rho = {r \over {\sqrt{r^2-1}}}~e^{-t}~~~~~~~~
y = {1 \over {\sqrt{r^2-1}}}~e^{-t}~~~~~~~~({\rm Region~I})}

In \cseven, the point $(y,\vx)$ is in region II while the points labelled
by $y'$ are all in region I.
Since the trajectory has $r = r' = 0$ we have
$y' = e^{-t'}$. For a point $(r,t,\theta_2)$ in region II the field is
\eqn\eseven{ \eqalign{ \Phi (r,t,\theta_2)
= \int_{-\infty}^{t + \log~{\tilde r}}dt'\int 
dk~k~d\theta~(e^{-t'}{1\over {\tilde r}}e^{-t})~
e^{ik{\tilde \rho} \cos\theta}~
[J_{-\nu} & ({|k|\over {\tilde r}}  e^{-t})J_{\nu}(|k|e^{-t'})\cr 
& - J_{-\nu}(|k|e^{-t'})J_{\nu}({|k|\over {\tilde r}}e^{-t})]}}
where we have defined
\eqn\eeight{ {\tilde r} = {\sqrt{r^2-1}}~~~~~~~
{\tilde \rho} = {r \over {\sqrt{r^2-1}}}~e^{-t}}
To determine the one point function in static coordinates we have to now
go to the boundary at $r = r_0 >> 1$. Then ${\tilde r} \sim r_0$ and
${\tilde \rho} \sim e^{-t}$. Peforming (in order) the angular integration over
$\theta$, the integral over $t'$ and finally the integral over $|k|$ exactly
as in Section 4 it is easy to see that
\eqn\enine{
< \cO_- (t,\theta_2)>_{static} = 
{\rm Lim}_{r_0 \rightarrow \infty} [(r_0)^{\nu - 1}~
\Phi(r,t,\theta_2)] \sim {\rm constant}}

Indeed the restriction of the coordinate transformations \esixa\ to
$\scri^+$ is precisely the conformal transformation \ethree, with the
identifications
\eqn\eten{w = t + i\theta_2}
This explains why the conformal transformation \ethree\ render the one
point function constant.

\subsec{Geodesics in $dS_3$ : Global boundary}

The story is similar in global coordinates. Now the cutoff boundary
is at $T = T_0 = {\pi \over 2} - \epsilon$ with $\epsilon << 1$. The
transformation between the global coordinates and planar coordinates
of regions I and II are given by directly comparing the formulae in
Section 2,
\eqn\fone{\eqalign{& y = {\cos T \over \sin T - \cos \theta_1} \cr
& \rho = {\sin \theta \over \sin T - \cos \theta_1}}}
The worldline of the geodesic is now $\theta_1' = \pi$. The one
point function on the boundary may be now calculated by evaluating
the field in \cseven\ by substituting \fone\ and finally performing
the limit $\epsilon \rightarrow 0$. The final result is
\eqn\ftwo{
< \cO_- (\theta_1,\theta_2)>_{global} = 
{\rm Lim}_{(T_0 \rightarrow {\pi \over 2})} 
[ (\cos~T_0)^{\nu - 1}~
\Phi(T, \theta_1 ,\theta_2)] \sim ({1 \over \sin \theta_1})^{1-\nu}}

To understand this result, consider the transformation between global
coordinates and static coordinates in Region II. These are
\eqn\fthree{r = \sec T~\sin \theta_1~~~~~~~~~
\tanh t = - {\rm cosec}~ T~\cos \theta_1}
On $\scri^+$ this becomes
\eqn\ffour{ \tanh t = -\cos \theta_1}
or equivalently
\eqn\ffive{ \tan {\theta_1 \over 2} = e^t}
In terms of standard complex coordinates on $S^2$
\eqn\fsix{ u = \tan {\theta_1 \over 2}~e^{i\theta_2}}
we therefore have the conformal transformation
\eqn\fseven{ u = e^w}
where $w$ has been defined in \eten. This is exactly the conformal
transformation between the planar complex coordinate $z$ and 
$w$. Thus one might have expected that on the sphere we should have
a one point function
\eqn\feight{ ({1 \over u \bu})^{1-\nu \over 2} = 
({1\over \tan {\theta_1 \over 2}})^{1-\nu}}
which is not the same as \ftwo. 

However there is a Weyl anomaly here since the metric on the sphere is
given in terms of $u,\bu$ by
\eqn\fnine{ds^2 = {4 du d\bu\over (1 + u\bu)^2}}
This means that while performing the conformal transformation we must
account for this conformal factor which is not a product of a function
of $u$ and a function of $\bu$. This is exactly what has to be done to
calculate two point functions of operators on the sphere. Taking this
into account we should get
\eqn\fine{ ({(1 + u\bu)^2 \over u \bu})^{1-\nu \over 2}}
which is precisely \ftwo.

\subsec{Higher dimensions}

Many of the above considerations would generalize to other dimensions
as well. For similar reasons, one point functions measured on the boundary
defined in terms of static coordinates would be a constant. For planar
and global coordinates the transformations to static coordinates are
in fact exactly the ones given above for all dimensions, and so would
be their restrictions to the boundary. The transformation laws for
one point functions would be however different.

For $AdS_{d+1}$ the boundary metric in static coordinates may be written as
\eqn\none{ds^2 = r_0^2[dt^2 + (1-\mu^2)d\phi^2 + {d\mu^2 \over 1-\mu^2}
+ \mu^2 d\Omega_{d-3}^2]}
The normal way to interpret this theory would be to consider $t$ as the
euclidean time.
The previous discussion suggests that there is another way to interpret
the theory, viz via the analytic continuation
\eqn\ntwo{\phi = i\eta}
and regarding $\eta$ as the time.
Then the metric \none\ becomes, 
\eqn\nthree{ds^2 = dt^2 - (1-\mu^2)d\eta^2 + {d\mu^2 \over 1-\mu^2}
+ \mu^2 d\Omega_{d-3}^2]}
This is the metric on $dS_{d-1} \times R$. Constant $\mu$ observers
will perceive the invariant vacuum as a thermal bath. In fact one
often uses the reverse of this argument to understand why there is
thermality in de Sitter spacetimes from the bulk viewpoint. It is
interesting that the holographic signature of thermality gets related
to the question of thermality in de Sitter spacetimes again, albeit
in two less dimensions.

\newsec{Conclusions}

We have offered a {\it signature} of the thermal properties of geodesic
observers in the holographic theory. This is not quite an {\it explanation}.
However we believe that this insight will be useful in a proper holographic
understanding of cosmological spacetimes. 

One important assumption in our work is that the one point function of
a CFT operator is given by the value of the bulk field on the
boundary, with suitable powers of the cutoff stripped off. We
calculated the field on the boundary in a standard fashion using
retarded Green's functions. This is entirely analogous to treatments
in the AdS/CFT correspondence. In the AdS/CFT correspondence this is
almost forced upon us since the bulk and the boundary share the same
``time'', and one would like to maintain causality. Things are less
clear in the present case. Here the boundary is either $\scri^+$ or
$\scri^-$. Our procedure gives the one point functions in the boundary
on $\scri^+$ and a zero value on $\scri^-$. Presumably to get one
point functions on $\scri^-$ one should use {\it advanced} Green's
functions in the bulk. Another possiblity is to use a symmetric
Green's function. In fact the latter is suggested by an interesting
result of \bms. These authors consider the one parameter class of
invariant vacua in the bulk \mot, or their planar coordinate analogs
\sv.  They show that a bulk geodesic
observer would detect particles in all these vacua.  However the
spectrum is thermal only for a preferred value of the parameter - the
one which leads to the analytic continuation of the euclidean bulk
vacuum. This result could be reconciled with our results if one uses a
symmetric Green's function to obtain the field in the presence of a
geodesic source, since while the retarded or advanced Green's function
does not depend on the vacuum parameter, the symmetric Green's
function does.

Our holograms are obtained from the fields produced by point sources.
In a natural extension of terminology, this is a ${1\over N}$ effect.
In the $AdS/CFT$ correspondence one needs nonlocal operators to probe
local physics in the bulk \ref\sussk{J. Polchinski, L. Susskind and
N. Toumbas, {\it Phys.Rev.} {\bf D60} (1999) 084006, 
{\tt hep-th/9903228}; N. Toumbas and L. Susskind, 
{\it Phys.Rev.} {\bf D61} (2000) 044001, 
{\tt hep-th/9909013}.}. Since we are probing the entire trajectory of
a particle one should have a better description in terms of these
objects and this would be a leading effect. It would be interesting
to see analogs of these in the $dS/CFT$ correspondence.

Finally, the presence of an external source in our discussion is not
quite natural. The proper formulation of the problem would be to 
consider a wavepacket made out of bulk fields whose center follows
a geodesic, as in \bulkboundary. The tails of these wavepackets
then provide the one point functions necessary for the hologram.
However one has to redo the analysis of thermal behavior by
``comoving'' set of observers in the bulk. Because of the nontrivial
profile of the wavepacket one would get rather complicated transformations
in the boundary theory. 

\newsec{Acknowledgements}

I would like to thank Sandip Trivedi for numerous discussions and
collaboration in the early stages of this work.  I also thank Samir
Mathur, Shiraz Minwalla and Al Shapere for discussions on various
aspects of this work and the referee for a helpful comment.  This work
is partially supported by DOE Grant No. DE-FG01-00ER45832.

\listrefs
\end